\documentstyle{mn}

\title{Determination of the dynamical structure of galaxies using
optical spectra} 

\author[S. De Rijcke \& H. Dejonghe]{S. De Rijcke \thanks{Research
Assistant of the Fund for Scientific Research - Flanders
(Belgium)(F.W.O)} and H. Dejonghe \\ 
Sterrenkundig Observatorium, University of Ghent, \\ 
Krijgslaan 281, S9, 9000 Gent, Belgium}

\begin{document}

\maketitle

\begin{abstract} 
Galaxy spectra are a rich source of kinematical information since the
shapes of the absorption lines reflect the movement of stars along the
line-of-sight. We present a technique to directly build a dynamical
model for a galaxy by fitting model spectra, calculated from a
dynamical model, to the observed galaxy spectra. Using synthetic
spectra from a known galaxy model we demonstrate that this technique
indeed recovers the essential dynamical characteristics of the galaxy
model. Moreover, the method allows a statistically meaningful error
analysis on the resulting dynamical quantities.
\end{abstract}

\begin{keywords}
Galaxies: kinematics and dynamics--Celestial mechanics, stellar dynamics
\end{keywords}

\section{Introduction}

The construction of a dynamical model for a galaxy depends on two
kinds of data. The photometric measurements yield, after deprojection,
the mass density. Assuming a mass-luminosity ratio, the Poisson
equation then leads to the gravitational potential. The kinematics are
obtained by analysing spectral information, with the aim to
characterise the line-of-sight velocity distribution (hereafter LOSVD)
$\phi(v_p,x_p,y_p)$, which is the probability of finding a star with
projected velocity $v_p$ at a position $(x_p,y_p)$ on the sky.

When the signal to noise ratio ($S/N$) of the data is rather low, a
LOSVD can be parametrised as a simple Gaussian. There exist a host of
methods to determine the best fitting Gaussian LOSVD, such as the
Fourier quotient method (Simkin 1974, Sargent et al 1977), modified by
Winsall \cite{winsall} to incorporate also non-Gaussian LOSVDs, the cross
correlation method (Tonry \& Davis 1979), the Fourier fitting method
(Franx, Illingworth \& Heckman 1989) and the Fourier correlation
quotient method (Bender 1990).

Motivated by the increasing quality of the data, new methods have been
developed recently that are able to quantify symmetric and
antisymmetric deviations of the LOSVD shape from a Gaussian (e.g. van
der Marel \& Franx 1993 and Gerhard 1993). A LOSVD is then
parametrised by the average projected velocity $\langle v_p \rangle$,
the projected dispersion $\sigma_p$ and the lowest order coefficients
of a Gauss-Hermite expansion of the LOSVD, $h_3$ and $h_4$. These can
subsequently be used in a fitting routine to construct a dynamical
model that best represents the kinematical data.

Another path leading to estimates of the shape of LOSVDs employs
non-parametric fitting procedures such as the Wiener filtering method
(Rix \& White 1992), the maximum entropy method (Statler 1991), the
unresolved Gaussian method (Kuijken \& Merrifield 1993) and the
Bayesian fitting method (Saha \& Williams 1994). These give a totally
unbiased estimate of the LOSVDs. 

It is clear that all these modeling techniques require two fits. One
fit is needed to extract the kinematics from the spectra, and a second
fit then produces a dynamical model. In this paper we propose a method
to construct a distribution function that operates directly on the
available spectra, after they have undergone all the necessary steps
of a standard data reduction. This problem is well defined, as is
shown by Dejonghe \& Merritt \cite{dejonghe:merritt}. The essentials
of the method are explained in the next section. In section 3, we test
the method on a set of spherical anisotropic Plummer models.

\section{The modeling procedure}

A galaxy spectrum, obtained at a given position $(x_p,y_p)$ on the
sky, is the integrated light of all the stars along the
line-of-sight. The spectrum of each contributing star is Doppler
shifted according to its velocity relative to the observer. If we
assume that the bulk of the light of the galaxy is produced by stars
of approximately the same spectral class (in the case of an elliptical
galaxy, most of the light originates from the G to M giants) we can
write the following form for a galaxy spectrum at $(x_p,y_p)$
\begin{eqnarray}
g(x_p,y_p,\lambda) &=& \int \phi(v_p,x_p,y_p) \, s(\lambda(v_p))\,dv_p 
\label{conv1}
\end{eqnarray}
which expresses the spectrum as a sum of products of probabilities :
the probability of finding a star with projected velocity $v_p$ at a
position $(x_p,y_p)$ and the probability that such a (standard) star
emits a photon with Doppler shifted wavelength $\lambda$, according to
its spectrum in the rest frame $s(\lambda)$.  We denote the rest
wavelength by $\lambda(v_p)$.  For velocities which are small compared
to the velocity of light one has the Doppler shift formula
\begin{equation}
\frac{v_p}{c} = \frac{\lambda - \lambda(v_p)}{\lambda(v_p)} \approx
\ln \frac{\lambda}{\lambda(v_p)}.
\end{equation}
Obviously, (\ref{conv1}) can be written in the form of a convolution 
using logarithmically rebinned spectra, which also eliminates the 
choice of a dimension for the wavelengths. Hence
\begin{equation}
g(x_p,y_p, \ln\lambda) = \int \phi(v_p,x_p,y_p) \, s(\ln \lambda-\frac{v_p}{c})\,dv_p. 
\label{1}
\end{equation}

It is always possible to write the distribution function (hereafter
DF) $F({\rm I})$, where ${\rm I}$ are the isolating integrals of the
motion, as an infinite sum of basis functions $F^i({\rm I})$ with
coefficients $c_i$
\begin{equation}
F({\rm I}) = c_i\,F^i({\rm I}).
\end{equation}
In the sequel, we will consistently use a summation convention.  These
basis functions are the DFs of simple dynamical models, called the
components, that form a complete set.  Denoting by $v_x$ and $v_y$ the
velocity components in the plane of the sky, and by $z$ the coordinate
along the line-of-sight, we find formally
\begin{equation}
\phi(v_p,x_p,y_p) = \int_{-\infty}^{\infty} dz \!\! \int \!\! \int 
F({\rm I}) \,dv_x\,dv_y.   \label{2}
\end{equation}
Obviously, the LOSVD is linear in the DF, and can also be written as a
weighed sum of component LOSVDs $\phi^i(v_p,x_p,y_p)$ with the same
coefficients $c_i$ as the DF. Likewise, we obtain component spectra 
\begin{equation}
g^i(x_p,y_p,\ln \lambda) = \int \phi^i(v_p,x_p,y_p) \, s(\ln \lambda-\frac{v_p}{c})\,dv_p.
\label{3}
\end{equation}
We label with the composite index $n$ the $k$-th pixel of the $l$-th
galaxy spectrum (we label each spectrum and its corresponding
line-of-sight with a number $l$) and denote its value by $g_n$. A
similar notation applies for the spectra of the $i$-th component,
$g_n^i$. It is easy to see that ideally
\begin{equation}
g_n = c_i\,g^i_n 
\hspace{1cm} n=1,\ldots,N,  \label{voeg1}
\end{equation}
with $N$ the total number of pixels considered.

Since we can only handle a truncated expansion of the DF, we will have
to use a limited number $m$ of components and try to find those
coefficients $c_i$ that give rise to a model that best represents the
data. This we do by using a Quadratic Programming minimisation
technique (Dejonghe 1989). The stellar template spectrum
is normalised to one \begin{equation} \int_{-\infty}^{\infty} s(\ln
\lambda)\,d \ln \lambda = 1, \label{refie} \end{equation} where the
function $s(\ln \lambda)$ is set to zero in the region outside the
data. The aim is to model the observed spectra with
(\ref{voeg1}). Since the component LOSVDs $\phi^i(v_p,x_p,y_p)$ are
normalised to their projected mass density $\rho_p^i(x_p,y_p)$, the
calculated component spectra $g^i(x_p,y_p,\ln \lambda)$ will also be normalised to
$\rho_p^i(x_p,y_p)$. In order to make (\ref{voeg1}) correct, in the
sense that
\begin{equation}
\rho(x_p,y_p) = c_i \, \rho_p^i(x_p,y_p), 
\end{equation} 
the observed galaxy spectra $g(x_p,y_p,\ln \lambda)$ must be
normalised to the projected mass density $\rho_p(x_p,y_p)$. However,
an observed galaxy spectrum cannot be written exactly as the
convolution of a template spectrum and a LOSVD since such a
convolution has two rapidly decaying wings in both regions where the
template spectrum is unknown and by default set to zero. Observed
galaxy spectra are also unknown and set to zero in the same regions as
the template spectra. So by normalising the galaxy spectra to the
projected mass density a small error is introduced, because the two
small but unknown wings, required to make the convolution formally
exact, are neglected.

Usually the continuum of a galaxy spectrum and the continuum of a
weighed sum of component spectra will not be the same. That is why the
continua of the galaxy spectra and the stellar template spectrum have
to be subtracted before they are normalised. For simplicity, we still
use the notation $g(x_p,y_p,\ln \lambda)$, now for a continuum
subtracted galaxy spectrum, and $s(\ln \lambda)$ for the continuum
subtracted stellar spectrum.

The quantity that has to be minimised is 
\begin{equation}
\chi^2 = w^n \left( g_n - c_i\,g_n^i \right)^2, \label{5}
\end{equation}
with $w^n=1/ \sigma_n^2$ and $\sigma_n$ the noise on $g_n$,
which is well modeled by Poisson noise. This expression 
can be expanded into the form
\begin{equation}
\chi^2 = c_i\,D^{ij} c_j -2 p^i c_i + e
\end{equation}
or, in matrix notation
\begin{equation}
\chi^2 = {\rm c}^t {\rm{\bf D}} {\rm c} - 2 {\rm p}^t {\rm c} + e. \label{4}
\end{equation}
The matrix ${\rm{\bf D}}$ with elements 
\begin{equation}
D^{ij} = w^n g^i_n g^j_n
\end{equation}
is called the Hessian matrix and the vector ${\rm p}$ has components
\begin{equation}
p^i = w^n g^i_n g_n.
\end{equation}
The scalar constant $e$ is given by
\begin{equation}
e = w^n g_n g_n. 
\end{equation}
This $\chi^2$ is quadratic in the coefficients. It has to be minimised
under the linear constraint that the DF has to be positive
\begin{equation}
c_i F^i({\rm I}) \geq 0.
\end{equation}
The positivity is tested on a grid in phase space. This is a typical
problem of Quadratic Programming. Moreover, the $\chi^2$ has the
statistical meaning of a goodness of fit so its expectation value
equals
\begin{equation}
\langle\chi^2\rangle = N-m-1.
\end{equation}
Other observational data can be added to this $\chi^2$. However, it 
looses its meaning as a goodness of fit when other weights than
the errors are used in (\ref{5}). In appendix C we present the error
analysis on the obtained coefficients.

\setcounter{figure}{0}
\begin{figure*}
\vspace{7cm}
\special{hscale=60 vscale=60 hsize=570 vsize=210 
	 hoffset=0 voffset=285 angle=-90 psfile="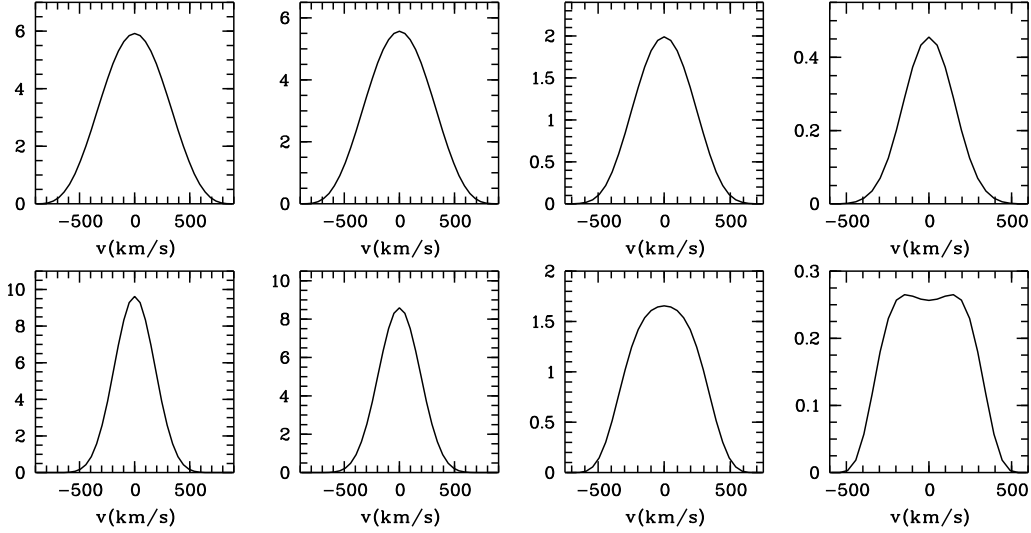"}
\caption{The LOSVDs of a radial Plummer model with $q=1$ (top) and a
tangential Plummer model with $q=-6$ (bottom), both with $M=5\times 
10^{11}M_\odot$ and $r_c = c = 5$kpc, for $r_p = 0$, $1$, $5$
and $10$kpc (from left to right). \label{fig14}}
\end{figure*}

\section{Putting the method to the test}

To test our modeling procedure we make use of synthetic galaxy spectra
with simulated Poisson noise. These are created with spherical
anisotropic Plummer models (Dejonghe 1987). The synthetic
spectra are then used as input for our modeling procedure. As
components we use generalised Fricke models.

\subsection{The Plummer model}

The Plummer potential can be written as 
\begin{equation} 
\psi(r) ={{GM} \over r_c} {1 \over \sqrt{1+(r/r_c)^2}} 
\end{equation} 
with $r_c$ the so-called core radius and $M$ the total mass.  We take
$\sqrt{\psi(0)} = \sqrt{GM/r_c}$ as the unit of velocity so the central value
of the potential becomes unity. The density of the anisotropic Plummer
models can be written as
\begin{equation} 
\rho(r) = \psi(r)^{5-q}(1+(r/c)^2)^{-q/2} 
\end{equation} 
with $q$ and $c$ real numbers. This function is an augmented mass
density that is instrumental in the construction of the 2-integral DF
$F(E,L)$ (Dejonghe 1986). Inserting the functional form
for the Plummer potential in this expression, with $r_c = c$, one sees
that all Plummer models have the same density distribution
\begin{equation} 
\rho(r) = (1+(r/c)^2)^{-5/2}.
\end{equation} 
We will approximate the anisotropic Plummer model with Fricke components with 
augmented mass density
\begin{equation}
\rho(r) = \psi(r)^\alpha (r/c)^{2 \beta} 
\end{equation} 
with $\alpha$ and $\beta$ real numbers. The characteristics of both
the Plummer models and the Fricke components can be calculated using a
more general expression for the mass density
\begin{equation} 
\rho(r) = \psi(r)^\alpha \left( \frac{r}{c} \right)^{2 \beta} \left(
1+ \left( \frac{r}{c} \right)^2 \right)^{-(\beta+\gamma)} \label{massd}
\end{equation} 
with $\alpha$, $\beta$ and $\gamma$ real numbers that satisfy
\begin{equation} \alpha > 3-2 \gamma
\end{equation} 
to keep the total mass finite. Choosing $\beta=0$, $\alpha=5-q$ and
$\gamma=q/2$ one recovers the density of the Plummer models. The
choice $\gamma=-\beta$ yields the density of the Fricke
components. The advantage of the above-mentioned family lies in
the fact that most of the kinematics can be calculated
analytically. This is for instance the case for the anisotropic
velocity moments (see Appendix A)
\begin{eqnarray} 
\mu_{2n,2m}(r) &=& {\cal C}(m,n) \psi(r)^{\alpha+m+n}
\frac{(r/c)^{2\beta}}{(1+(r/c)^2)^{\beta+\gamma}} \times \nonumber \\
&& _2F_1(-m,\gamma+\beta;\beta+1,\frac{(r/c)^2}{1+(r/c)^2})
\end{eqnarray} 
with 
\begin{equation}
{\cal C}(m,n) =
\frac{2^{m+n}}{\sqrt{\pi}} \frac{\Gamma(n+1/2) \Gamma(\alpha+1) \Gamma(\beta+m+1)}
{\Gamma(\beta+1) \Gamma(\alpha+m+n+1)}.
\end{equation}
This general expression yields the following expressions for the
radial and tangential velocity dispersions,
\begin{eqnarray}
\rho(r) \sigma_r^2(r) \!\! &=& \!\! \mu_{2,0}(r) = \frac{1}{1+\alpha}
\psi(r)^{1+\alpha} \frac{(r/c)^{2 \beta}}{(1+(r/c)^2)^{\beta +
\gamma}} \\ 
\rho(r) \sigma_\phi^2(r) \!\! &=& \!\! \rho(r) \sigma_\theta^2(r) =
{1 \over 2} \mu_{0,2}(r) = \frac{1+\beta}{1+\alpha} \psi(r)^{1+\alpha}
\times \nonumber \\ 
&& \hspace{-1cm} \frac{(r/c)^{2
\beta}}{(1+(r/c)^2)^{\beta + \gamma}} \left( 1 - \frac{\gamma +
\beta}{1+\beta} \frac{(r/c)^2}{(1+(r/c)^2)} \right).
\end{eqnarray}
Binney's anisotropy parameter then reads
\begin{equation} 
\beta(r) = 1- \frac{\sigma^2_\phi}{\sigma^2_r}= \frac{-\beta+\gamma (r/c)^2}{1+(r/c)^2}.
\end{equation} 
For a detailed derivation of the anisotropic velocity moments, we
refer to appendix A.

One easily sees that the Fricke components have a constant anisotropy
\begin{equation} 
\beta(r)=-\beta.  
\end{equation}
The tangentially anisotropic components have zero central density
whereas the radially anisotropic ones have a central density cusp. The
isotropic components have a nonzero and finite central density. For
the Plummer models
\begin{equation}
\beta(r) = \frac{q}{2}\frac{(r/c)^2}{1+(r/c)^2}.
\end{equation}
We see that $\beta(r)$ and $q$ have the same sign and since $-\infty <
\beta(r) < 1$ the allowed range for $q$ is $-\infty < q < 2$. The true
nature of the orbital structure of a model becomes clearly visible in
the kinematics at radii $r \geq c$.

A LOSVD of a tangential Plummer model along a line-of-sight close to
the center is narrow and peaked. Moving the line-of-sight outward, the
LOSVDs become progressively broader and ultimately obtain a bimodal
shape. The LOSVDs of a radial model, on the contrary, will be broader
towards the center and narrower at large projected radii. As an
illustration, Fig. \ref{fig14} shows the LOSVDs of a radial Plummer
model with $q=1$ (top) and a tangential Plummer model with $q=-6$
(bottom) for $r_p = 0$, $1$, $5$ and $10$kpc.

\subsection{The synthetic spectra}

We want to construct spectra with (\ref{1}) so it is important that we
can calculate the necessary LOSVDs efficiently. It turns out that for
a subset of the models one can perform the integrations of the DF over
the velocity components in the plane of the sky analytically. This
yields $lp(r,r_p,v_p)$, the probability of finding a star with
projected velocity $v_p$ at a distance $r$ from the center of the
cluster and on a line-of-sight at a distance $r_p$ in the plane of the
sky from the center. The following expression is valid only when
$\beta$ is a positive integer and $\beta+\gamma$ a negative integer
(see Appendix B) :
\begin{eqnarray} 
lp(r,r_p,v_p) &=& \frac{1}{\sqrt{2\pi}}
\frac{\Gamma(\alpha+1)}{\Gamma(\alpha+ 1/2)}
\frac{(r/c)^{2\beta}}{(1+(r/c)^2)^{\beta+\gamma}} \times \nonumber \\
&& \hspace{-1.5cm} \psi(r)^{\alpha-1/2} \sum_{j=0}^{-(\beta+\gamma)}
(\beta+\gamma)_j \frac{\xi^j}{j!} (\sin \eta)^{2j} \times \nonumber \\
&& \hspace{-1.5cm} \sum_{i=0}^\beta \left(\frac{1}{2}\right)_{i+j}
\frac{(-\beta)_i}{(i+j)!}
\left(1-\frac{v_p^2}{2\psi(r)}\right)^{(-(i+j)+\alpha-1/2)} \times
\nonumber \\ 
&& \hspace{-1.5cm} \frac{(\sin \eta)^{2i}}{i!}
\,_2F_1(-(i+j),\alpha;\frac{1}{2};\frac{v_p^2}{2\psi(r)}).
\end{eqnarray}
Here 
\begin{equation} 
\xi = \frac{(r/c)^2}{1+(r/c)^2}
\end{equation}
and $\eta(r,r_p)$ is the angle between the spherical radial direction
and the line-of-sight at a certain point of the line-of-sight. The
integration along the line-of-sight has to be performed
numerically. The LOSVDs of the models we use are particularly
simple. In the case of Fricke components the above expression is
reduced to a sum over $i$, for Plummer models a sum over $j$
remains. LOSVDs of models that do not have positive integer $\beta$ or
negative integer $\beta + \gamma$ have to be calculated numerically
using (\ref{2}).

We calculate spectra out to about two effective radii. For a de
Vaucouleurs law, a distance of two effective radii encloses
approximately 69\% of the total luminosity. For a Plummer galaxy this
corresponds to $r_p = 1.49 c$.

Besides a LOSVD, one also needs a stellar template spectrum to
construct a galaxy spectrum. We use the spectrum of the K giant HR2425
in a wavelength range from 3900\AA$ $ to 5400\AA$ $. This spectrum is
rebinned logarithmically so that one pixel corresponds to a velocity
interval of 52 km/s. The Gaussian width ($\sigma_{\rm instr}$) of a line
that is intrinsically a delta function is about 150 km/s. The
synthetic spectra are calculated with (\ref{1}) and noise is added
according to a specified $S/N$ ratio.

\setcounter{figure}{1}
\begin{figure*}
\vspace{11cm}
\special{ hscale=68 vscale=70 hsize=570 vsize=300 
	 hoffset=-30 voffset=350 angle=-90 psfile="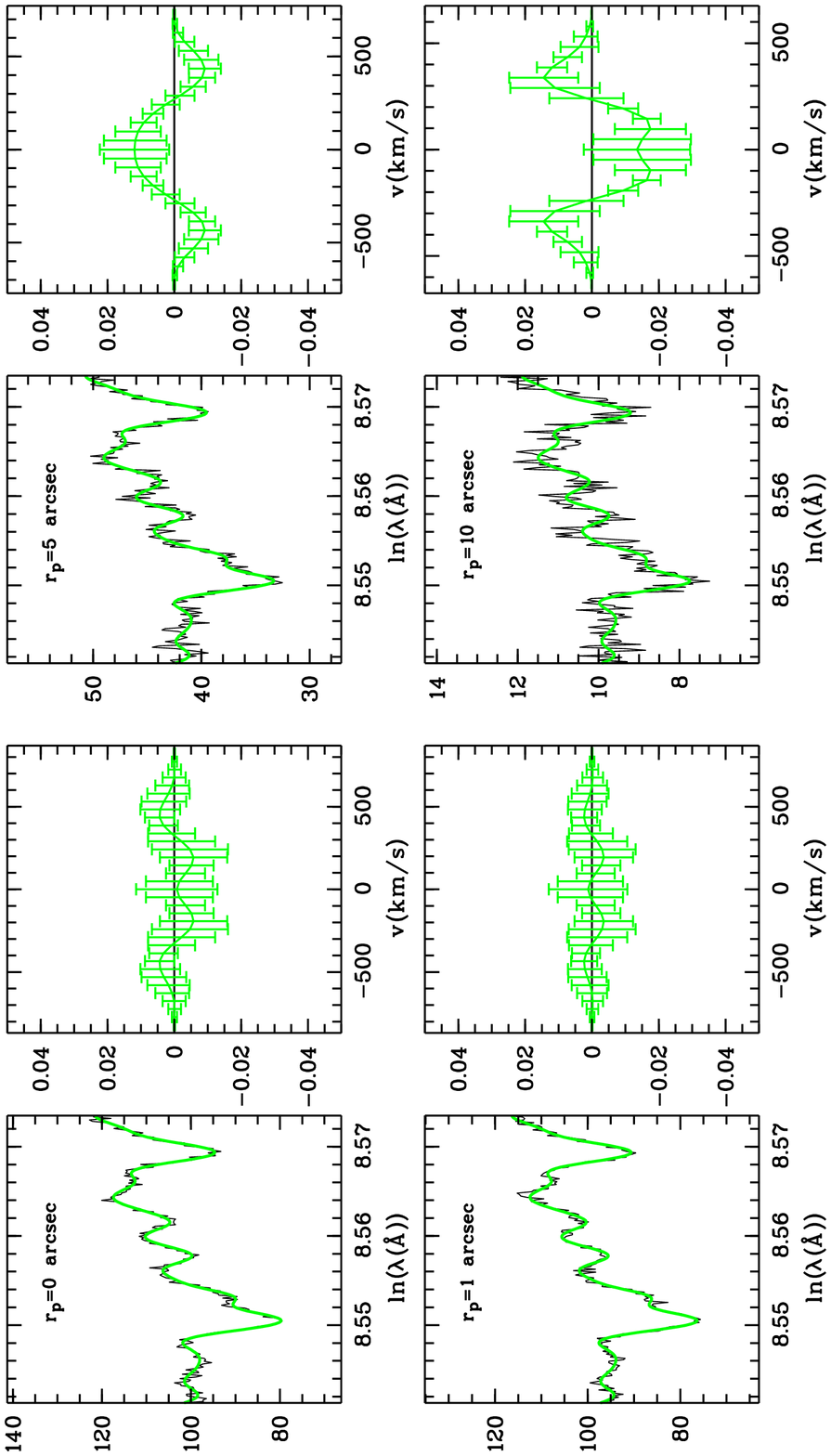"}
\caption{The synthetic spectra of a tangential $q=-$2 Plummer galaxy
(black) with central $S/N \approx 80$  for four different
lines-of-sight. The scaling of the ordinate is arbitrary. Over-plotted
in grey is the best fitting spectrum (the Fricke components with
$\alpha =4$, 5, 6, 8, 9, 10 and 12 are used). To the right of each
spectrum, we plotted ($\phi$(theo)-$\phi$(fit))/$\phi$(theo)$_{max}$
with $\phi$(theo) the Plummer LOSVD, $\phi$(fit) the fitted LOSVD and
$\phi$(theo)$_{max}$ the top value of the Plummer LOSVD. One can see
that the deviations between the Plummer LOSVD and the fitted LOSVD are 
nowhere larger than 1 to 2 \% of the top value.
\label{fig1}}
\setcounter{figure}{2}
\vspace{9cm}
\special{ hscale=60 vscale=50 hsize=560 vsize=260 
	 hoffset=0 voffset=280 angle=-90 psfile="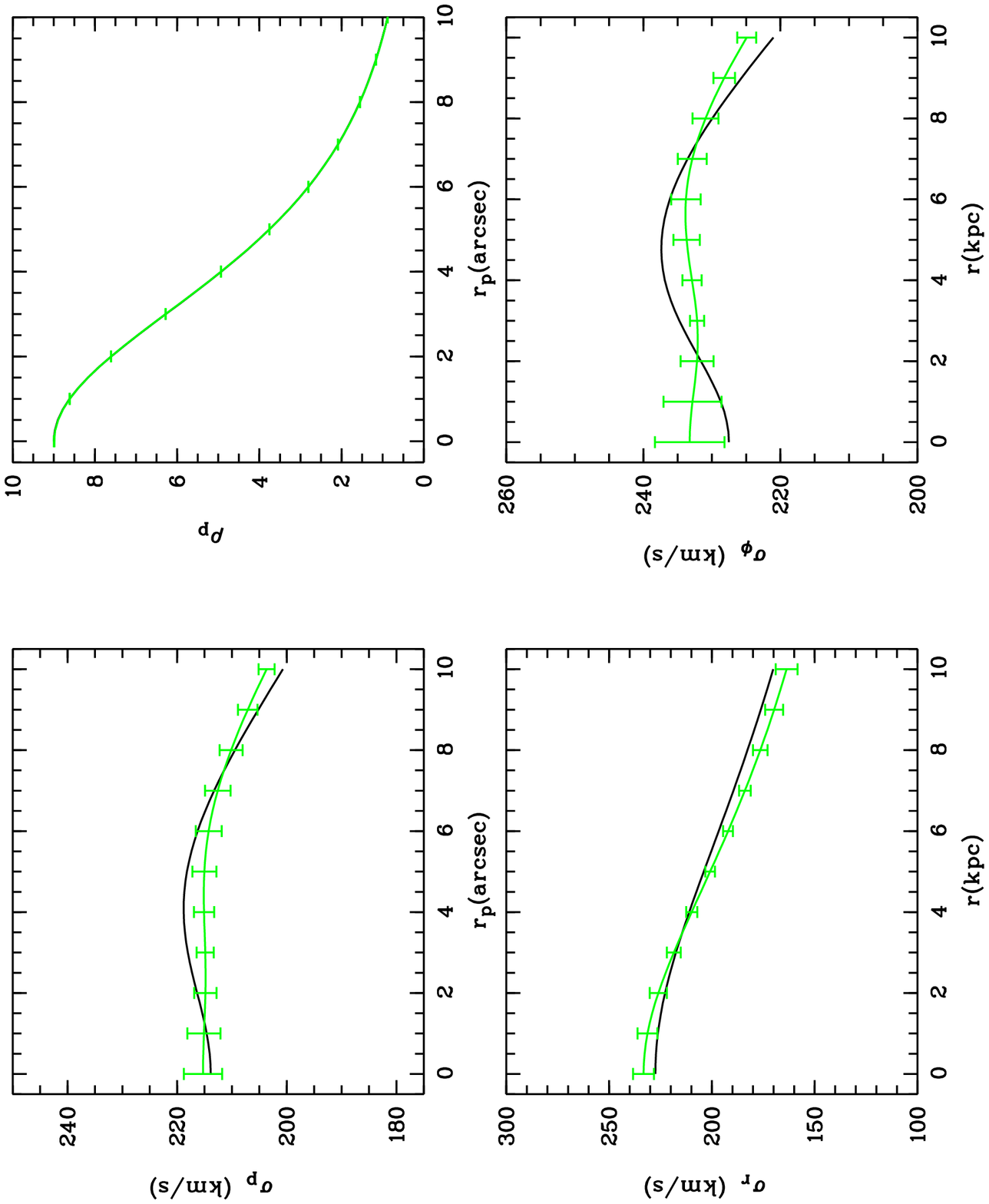"}
\caption{The kinematics of a tangential $q=-$2 Plummer galaxy
(black). Clockwise are plotted the projected velocity dispersion
$\sigma_p(r_p)$, the projected mass density $\rho_p(r_p)$ (in
dimensionless units), the tangential velocity dispersion
$\sigma_\phi(r) = \sigma_\theta(r)$ and the radial velocity dispersion
$\sigma_r(r)$. Shown in grey are the kinematics derived from the best
fitting model (using data with central $S/N \approx 80$  and Fricke
components with $\alpha =4$, 5, 6, 8, 9, 10 and 12). The deviations
are of the order of 5 km/s. The errorbars give a fair idea of the
actual errors on these derived quantities. There is one active
constraint. \label{fig2}}
\end{figure*}

\subsection{Testing the method}

To test if our modeling procedure can successfully perform the
inversion of the integral equation (\ref{1}) we offer the program
noisy synthetic spectra of a Plummer galaxy. We assume that the
Plummer galaxy contains only stars of the same spectral class. In that
case it is not necessary to subtract the continuum of the spectra. We
also use the stellar spectrum with which the synthetic spectra are
created as a template spectrum. These assumptions are needed to avoid
problems with bad continuum subtraction, template mismatch etc. that
have only to do with the data reduction and are common to all
methods. Due to the normalisation of the template spectrum and the
LOSVDs, the synthetic spectra will have automatically the proper
normalisation (see the discussion about the normalisation of spectra
in the paragraph below equation (\ref{refie})).

\subsubsection{A tangential $q=-2$ Plummer galaxy}

We first consider a mildly tangential $q=-$2 Plummer model. The model
has a core radius $r_c$ = $c$ = 6.75kpc and a total mass of 6.5$\times
10^{11} M_\odot$. This galaxy is placed at a distance of 206265 kpc so
that one arc-second ($''$) on the sky corresponds to a distance of 1
kpc. Two effective radii correspond to approximately 10$''$ so we
generate spectra for $r_p=0''$, $0.5''$, $1''$, \ldots,$10''$. We only
use a wavelength interval of about 170\AA$ $ containing the $M\!g${\sc
i} lines around 5180\AA$ $ and a strong $F\!e${\sc ii} line at
5270\AA$ $. During the fit, the correct mass and Plummer potential are
adopted.

The augmented mass density of this Plummer model can be written as a
binomial expansion, which shows that it can in principle be fitted
exactly in terms of Fricke components with $\alpha=7$. We first offer
the routine noiseless data. A library is set up that contains the
Fricke components with $\alpha=4$, 5, 6, 7, 8, 10 and 12 and with
positive integer $\beta$. These components are chosen because they
span a wide range in dynamical behavior. The models with high
$\alpha$-values are more centrally concentrated. The higher the value
of $\beta$ the more tangentially anisotropic the component
becomes. When dealing with real data, one has to make a reasonable
choice of components. In the following we will only use Fricke
components with positive integer $\beta$. This is just to save
calculation-time : the LOSVDs of these components can be calculated
analytically.

The 'redundant' components do not contribute significantly to the
result (they are not chosen or else given a very small coefficient)
and the correct components have the right coefficients. Thus, the
input model can be successfully recovered, which is of course a
prerequisite.

\setcounter{figure}{3}
\begin{figure*} 
\vspace{8cm}
\special{hscale=52 vscale=52 hsize=570 vsize=230 
	 hoffset=-13 voffset=270 angle=-90 psfile="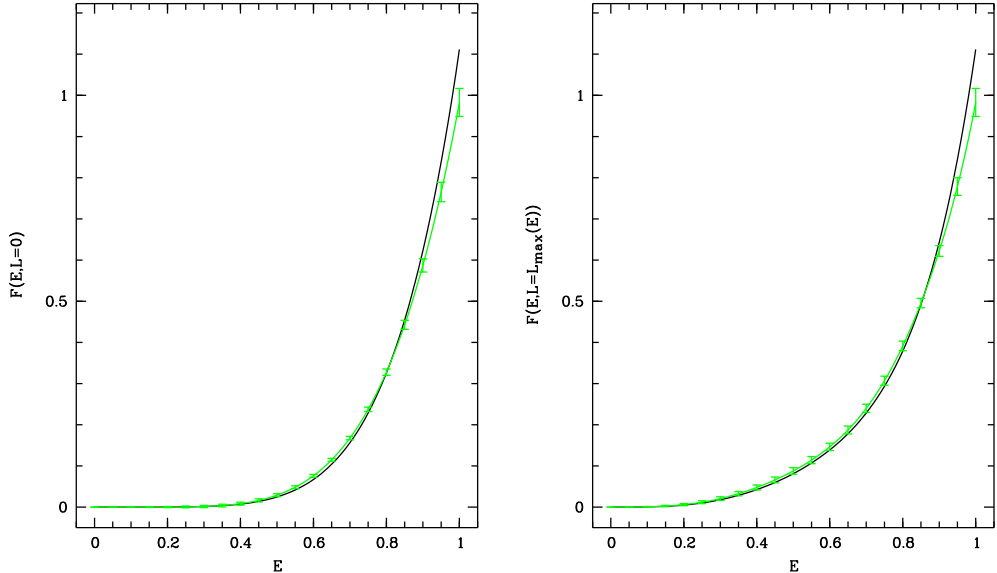"}
\caption{The DF of a $q=-2$ Plummer galaxy (black) and the fitted DF
(grey), with errorbars, shown for $L=0$ (left) and $L=L_{max}(E)$
(right). The fitted DF results from a fit to data with maximum $S/N
\approx 80$. Fricke components with $\alpha =4$, 5, 6, 8, 9, 10 and
12 are used. There is only one active constraint.
\label{fig10}}
\end{figure*}

We now consider data with a signal to noise ratio of $S/N \approx 80$
in the center, dropping to $S/N \approx 25$ at two effective
radii. Here and in the following, all signal to noise values are
expressed per pixel. The Fricke components with $\alpha=4$, 5, 6, 7,
8, 10 and 12 again serve as components. It appears that the spectra
constrain the LOSVDs to a high degree. The routine still uses the
correct components and assigns to them approximately the correct
coefficients but now also uses a few other components to reach a
minimum $\chi^2$. A more objective test is to fit the data without
incorporating the components that can represent the Plummer model
exactly in the library. The spectra are now fitted with a library
containing the Fricke components with $\alpha=4$, 5, 6, 8, 9, 10 and
12. This also gives an accurate approximation of the LOSVDs (see
Fig. \ref{fig1}). Although they are not included in the fit, the
observables $\sigma_p(r_p)$ and $\rho_p(r_p)$ are very well recovered
(see Fig. \ref{fig2}).

One can also use the errors on the coefficients to estimate the
uncertainty on the DF itself. Fig. \ref{fig10} shows the DF of the
$q=-2$ Plummer model and the DF that results of the previous fit. The
left plot shows a cut through the DF at $L=0$, the radial orbits. The
right one displays a cut along $L=L_{max}(E)$, the maximum angular
momentum an orbit with a certain energy can have, which is the locus
of the circular orbits. It can be seen that the fitted DF closely
follows the Plummer DF. Only in the center, where the DF is rather
peaked, is there a noticeable difference.

When spectra are used with $S/N \approx 55$ in the central regions,
dropping to $S/N \approx 15$ at two effective radii, the method still
yields a good approximation to the Plummer model (see Fig \ref{fig21}
and \ref{fig22}). Fig. \ref{fig20} compares the Plummer DF with the
fitted DF. There are no active constraints. The deviations between the
kinematics of the fitted model and of the Plummer galaxy are larger
but the errorbars, indicating the uncertainty on the coefficients, are
accordingly larger.

The analysis of about 20 spectra, each spanning a spectral region of
$\approx$ 200\AA$ $, with the limited number of components used
here, takes only a few hours on a Pentium {\sc ii} (this is the time
needed to calculate all the component spectra and to do the fit).

\setcounter{figure}{4}
\begin{figure*} 
\vspace{11cm}
\special{ hscale=68 vscale=70 hsize=570 vsize=300 
	 hoffset=-30 voffset=350 angle=-90 psfile="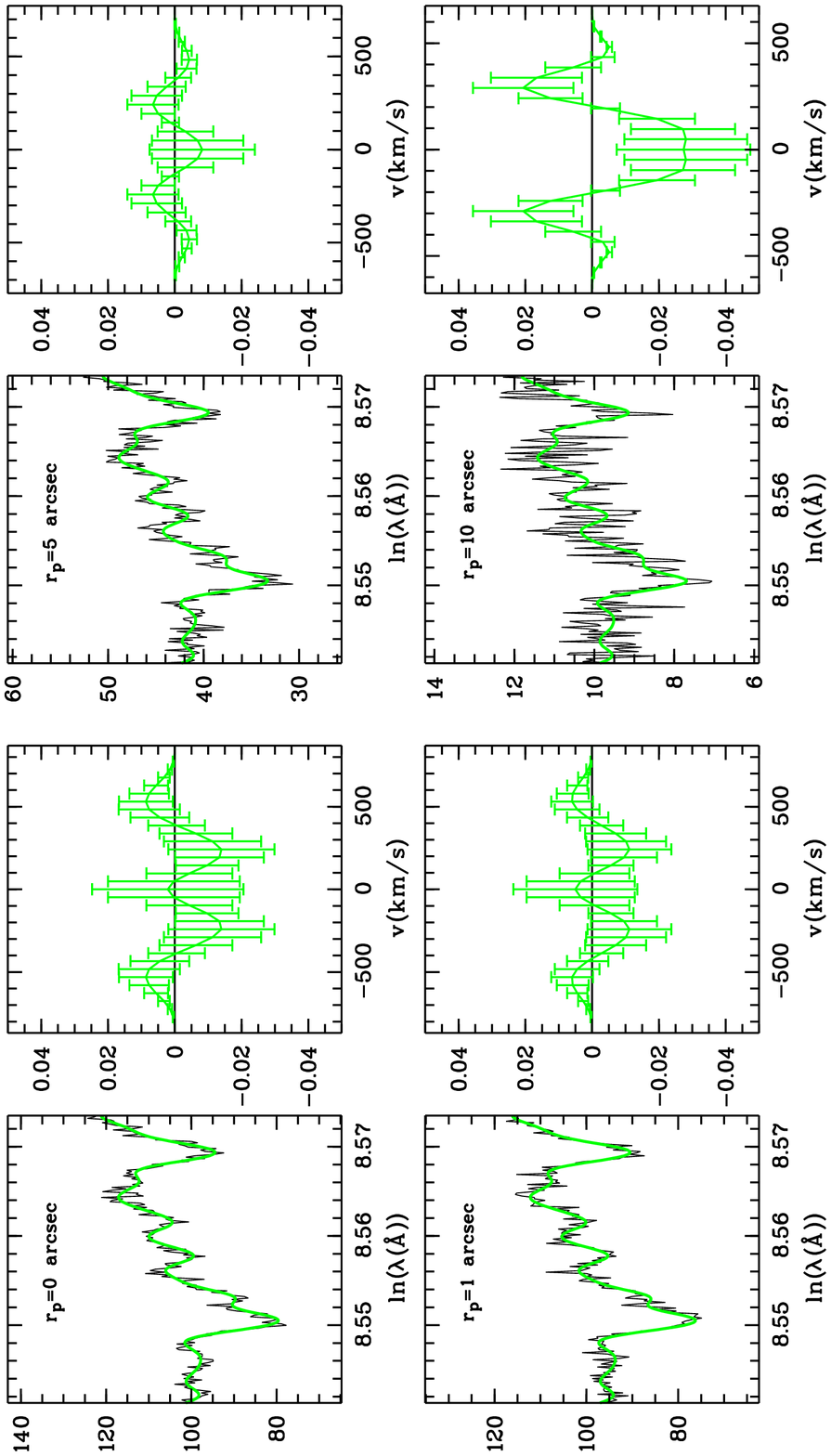"}
\caption{The same as Fig. \ref{fig1}
but for a fit to data with central $S/N \approx 55$, using the same Fricke
components. There are no active constraints.\label{fig21}}
\setcounter{figure}{5}
\vspace{9cm}
\special{ hscale=60 vscale=50 hsize=560 vsize=260
	 hoffset=0 voffset=280 angle=-90 psfile="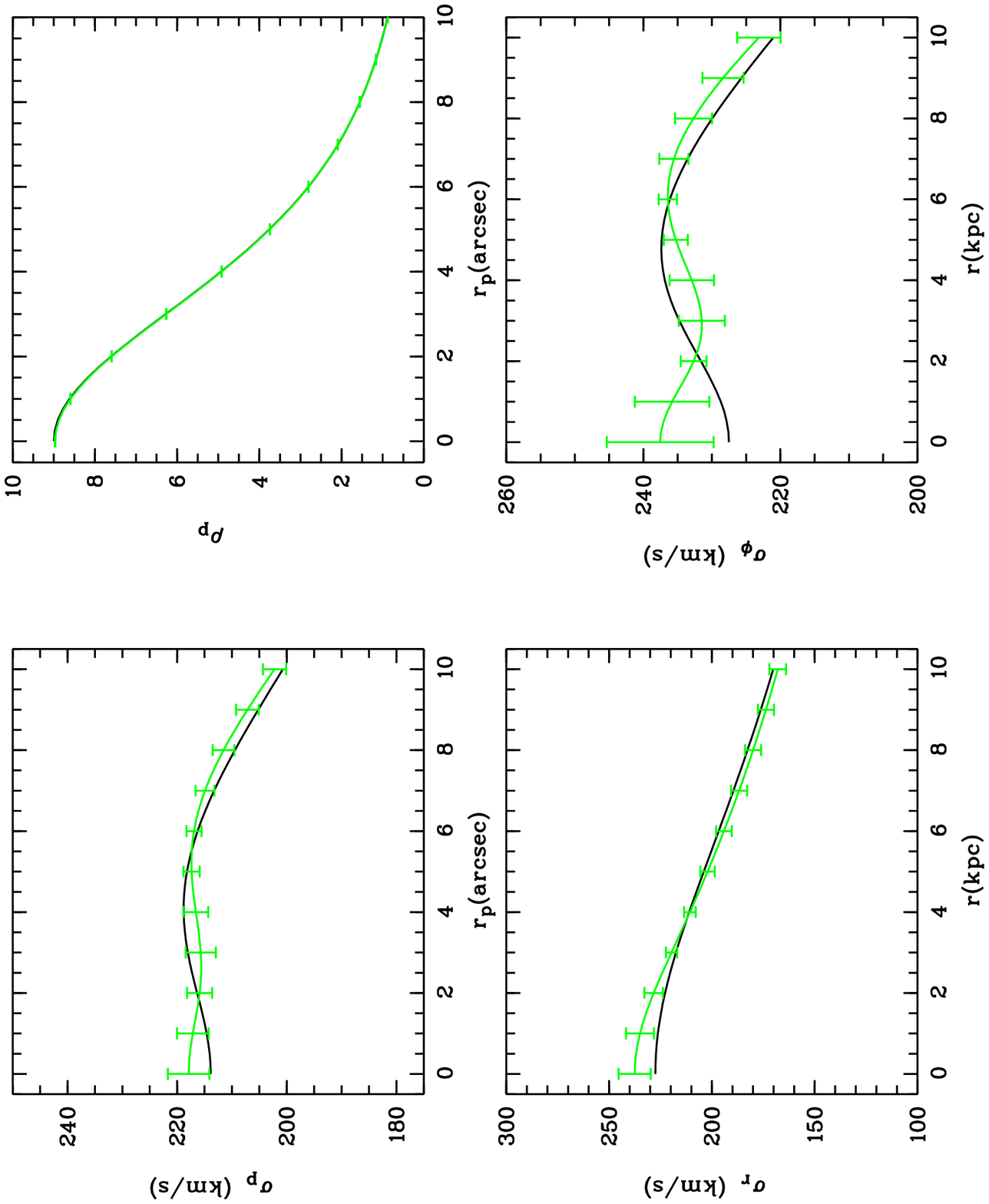"}
\caption{The same as Fig. \ref{fig2} but for a fit to data with maximum $S/N
\approx 55$, using the same Fricke components. There are no active
constraints.
\label{fig22}}
\end{figure*}

\subsubsection{Testing the errorbars}

To test if the errorbars give a reliable idea of the 1 $\sigma$
uncertainty on the coefficients, we construct 40 sets of 21 synthetic
spectra with $r_p=0''$, $0.5''$, $1''$, \ldots, $10''$ and a $S/N
\approx 80$ at $r_p$=0$''$. The spectra are generated with the $q=-2$
Plummer model of the previous paragraph.

\setcounter{figure}{6}
\begin{figure*}  
\vspace{8cm} 
\special{hscale=52 vscale=52 hsize=570 vsize=230 
	 hoffset=-13 voffset=270 angle=-90 psfile="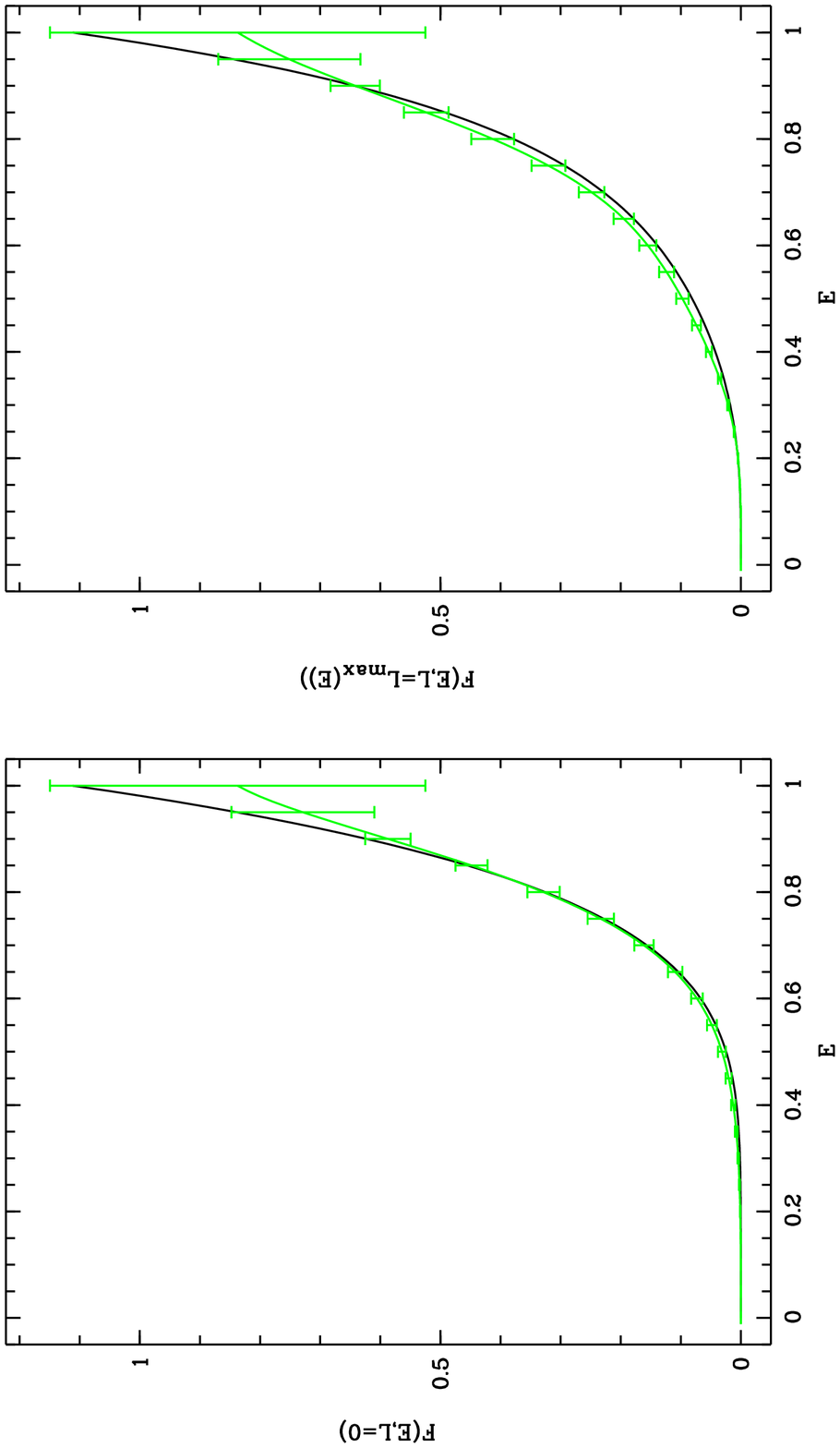"}
\caption{The same as Fig. \ref{fig10} but for a fit to data with maximum
$S/N \approx 55$.  \label{fig20}}
\end{figure*}

To each of these sets a model is fitted, using the Fricke components
$\alpha=4$, 6, 7, 8 and 10 with positive integer $\beta$ and assuming
the correct Plummer potential and mass. We then calculate for
$r_p=0''$, 1$''$, 5$''$ and 10$''$ and for selected $v_p$ the mean and
the dispersion (this we call the experimental dispersion) of the
LOSVDs of those 40 models. The average LOSVDs tend towards the true
LOSVDs and the 'experimental' and the theoretical $1 \sigma$-errorbars
essentially coincide, as can be seen in Fig. \ref{fig3}. Hence, the
errorbars are a reliable measure for the possible spread between the
input model and a typical fit, as caused by the adopted Poisson
noise. One also concludes that if the used component library is
sufficiently complete the systematic errors are negligible, compared
to the random errors caused by the Poisson noise.

\setcounter{figure}{7}
\begin{figure*} 
\vspace{14cm} 
\special{ hscale=75 vscale=75 hsize=570 vsize=450 
	 hoffset=-40 voffset=430 angle=-90 psfile="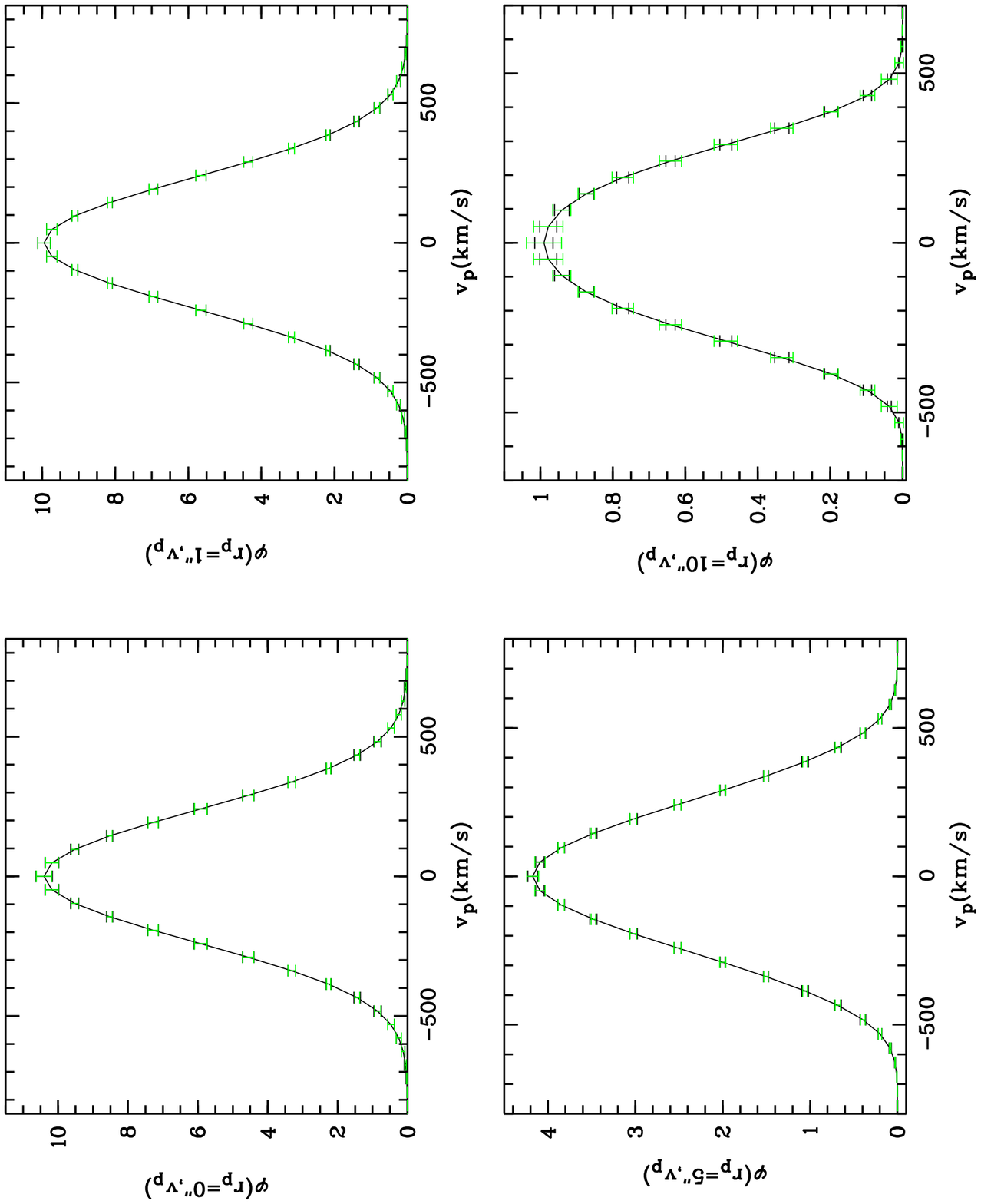"}
\caption{The average LOSVDs for $r_p=$0$''$, 1$''$, 5$''$ and
10$''$. Experimental (black) and theoretical (grey) 1 $\sigma$
errorbars are superposed. Where the errorbars coincide, the grey ones
hide the black ones. Three positivity constraints are
active. \label{fig3}}
\end{figure*}

\subsubsection{A tangential $q=-6$ Plummer galaxy}

This is a fairly tangential model. With the choice $r_c$ = $c$ = 5
kpc, the bimodal structure will reveal itself at $r_p \geq$ 5$''$. The
mass is $5 \times 10^{11} M_\odot$. Two effective radii now correspond
to 7.5$''$. Spectra are generated for $r_p=0''$, $0.5''$, $1''$,
\ldots, $7.5''$, once with a central $S/N \approx 80$ and once with a
maximum $S/N \approx 55$. This Plummer model can be written exactly as
a weighed sum of the four Fricke components with $\alpha=11$.

The Fricke models with $\alpha=7$, 9, 10, 11, 12, 13 and 14 serve as
components. The routine does not recognise the Fricke components with
$\alpha=11$ as the correct ones. They appear in the weighed sum with a
coefficient that is totally different from the one that would be
expected from the binomial expansion of the augmented mass density.
It is possible to obtain a very good fit to the data (see
Fig. \ref{fig5} and Fig. \ref{fig6} for data with a central $S/N
\approx 80$). The LOSVDs (including the bimodal shape at large radii)
and the kinematics are very well reproduced. Fig. \ref{fig30} compares
the Plummer DF and the fitted DF. There are no active constraints.

\setcounter{figure}{8}
\begin{figure*} 
\vspace{11cm}
\special{ hscale=68 vscale=70 hsize=570 vsize=300 
	 hoffset=-30 voffset=350 angle=-90 psfile="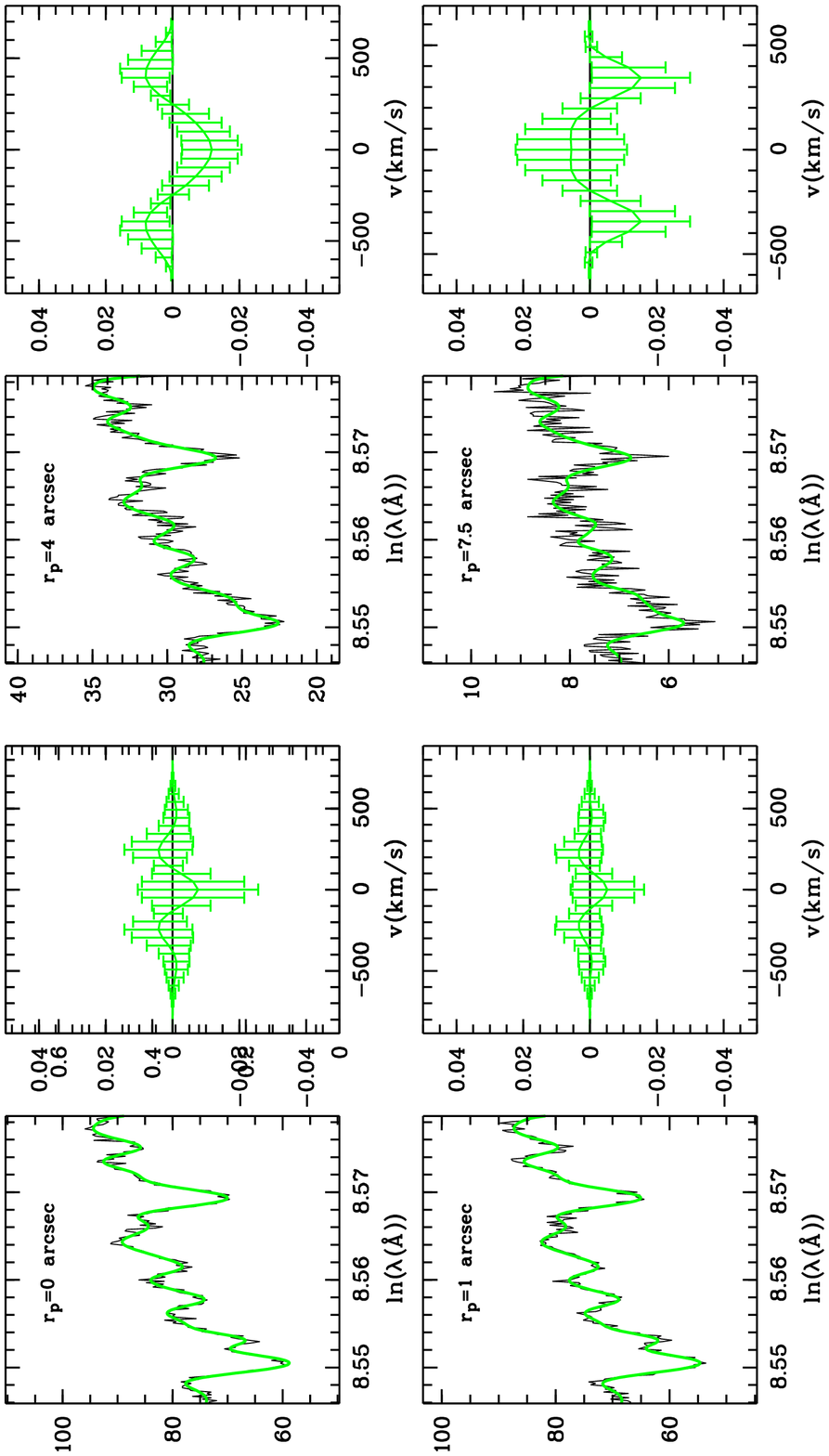"}
\caption{The same as in Fig. \ref{fig1}, but now for a tangential
$q=-$6 Plummer galaxy. The Fricke components with $\alpha=
7$, 9, 10, 11, 12, 13 and 14 are used in a fit to data with $S/N 
\approx 80$ for $r_p = 0''$. \label{fig5}}
\setcounter{figure}{9}
\vspace{9cm}
\special{ hscale=60 vscale=50 hsize=560 vsize=260 hoffset=0
          voffset=280 angle=-90 psfile="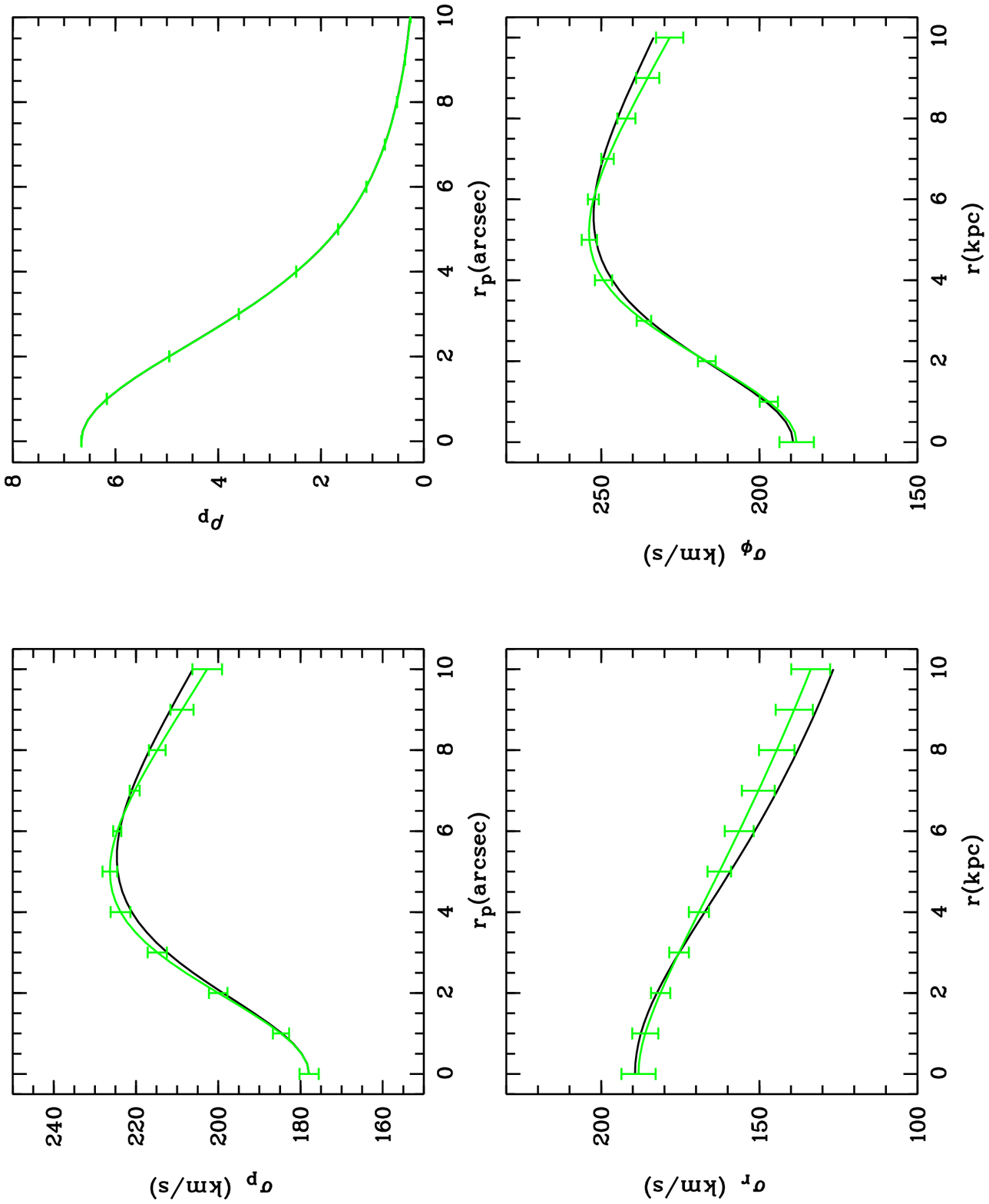"}
\caption{The same as in Fig. \ref{fig2}, but now for data from a tangential
$q=-$6 Plummer galaxy with maximum $S/N \approx 80$. \label{fig6}}
\end{figure*}

\subsubsection{A radial $q=1$ Plummer galaxy}

A radial model has much broader LOSVDs in the center than a tangential
model with the same mass. In the outer regions the inverse is true. As
a consequence, the outer spectra of a radial model still contain some
details that are totally smoothed out in the tangential
models. Therefore the inversion of the convolution integral (\ref{1})
is better behaved and the LOSVDs of a radial model are better
constrained by the data at large $r_p$ then those of a tangential
one. This Plummer model has $r_c$ = $c$ = 5 kpc and a total mass of $5
\times 10^{11} M_\odot$. Spectra are generated for $r_p=0''$, $0.5''$,
$1''$, \ldots, $7.5''$, once with a central $S/N \approx 80$ and once
with $S/N \approx 55$. As components we use Fricke components with
$\alpha=4$, 5, 6, 7 and 8.

\setcounter{figure}{10}
\begin{figure*} 
\vspace{8cm} 
\special{hscale=52 vscale=52 hsize=570 vsize=230 
	 hoffset=-13 voffset=270 angle=-90 psfile="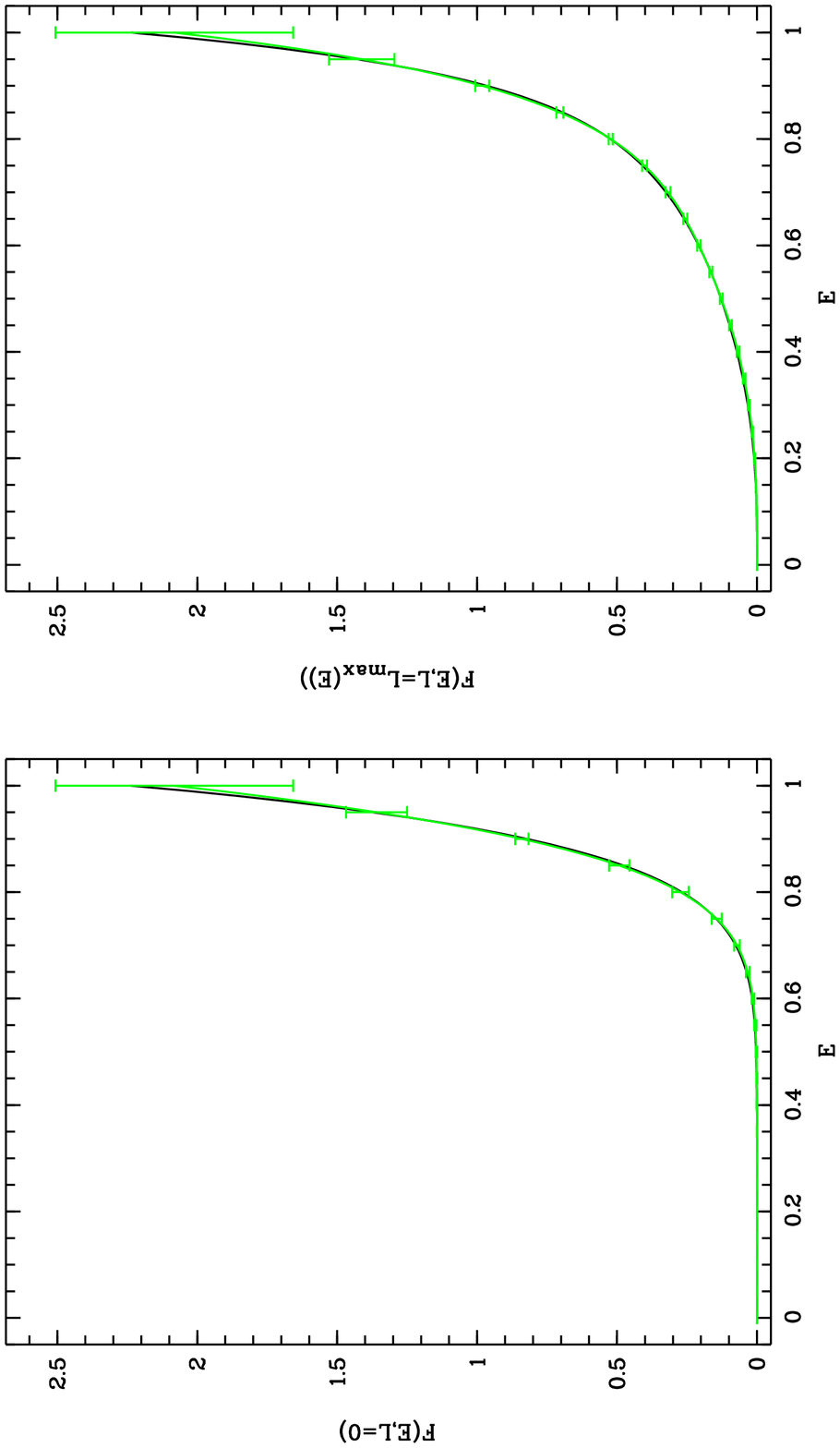"}
\caption{The same as Fig. \ref{fig10} but for a tangential $q=-6$ 
Plummer galaxy. \label{fig30}}
\end{figure*}

The LOSVDs of this model cannot be written exactly as a weighed sum of
Fricke components. Nonetheless, the LOSVDs are recovered very well in
both cases (see Fig. \ref{fig7} and \ref{fig8}). This is remarkable,
because all components are either isotropic or tangential. The only
way to construct a radial model with these components is to give
tangential components a negative coefficient to suppress the number of
stars on quasi-circular orbits.

Fig. \ref{fig40} compares the Plummer DF and the fitted DF and shows
the estimated $1 \sigma$ deviation between both DFs. 

\setcounter{figure}{11}
\begin{figure*} 
\vspace{11cm}
\special{ hscale=68 vscale=70 hsize=570 vsize=300 
	 hoffset=-30 voffset=350 angle=-90 psfile="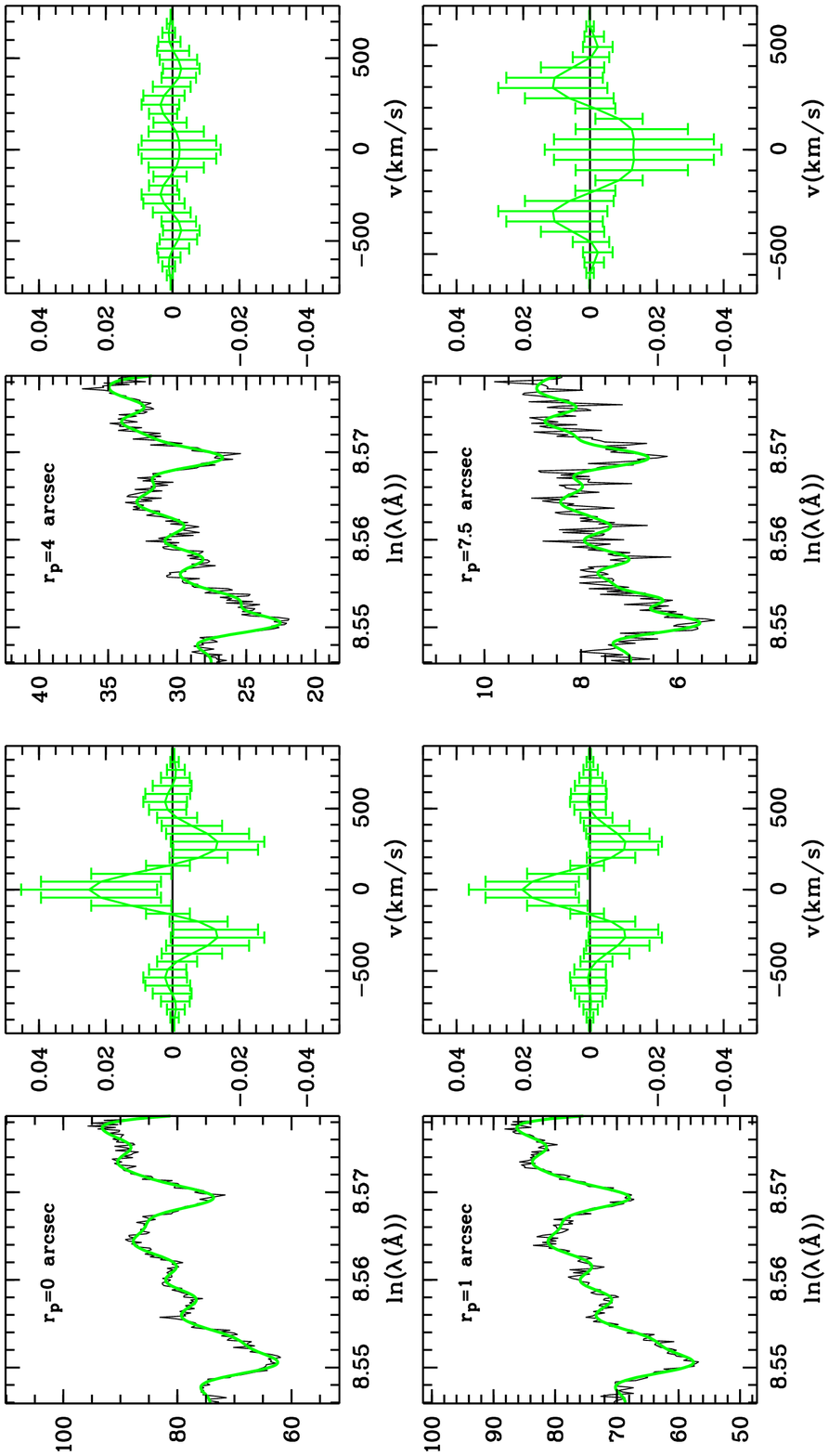"}
\caption{The same as in Fig. \ref{fig1}, but now for data from a
radial $q=$1 Plummer galaxy with maximum $S/N \approx 80$. We used
Fricke components with $\alpha=4$, 5, 6, 7 and 8. There are no active
constraints. \label{fig7}}
\setcounter{figure}{12}
\vspace{9cm} 
\special{hscale=60 vscale=50 hsize=560 vsize=260
	 hoffset=0 voffset=280 angle=-90 psfile="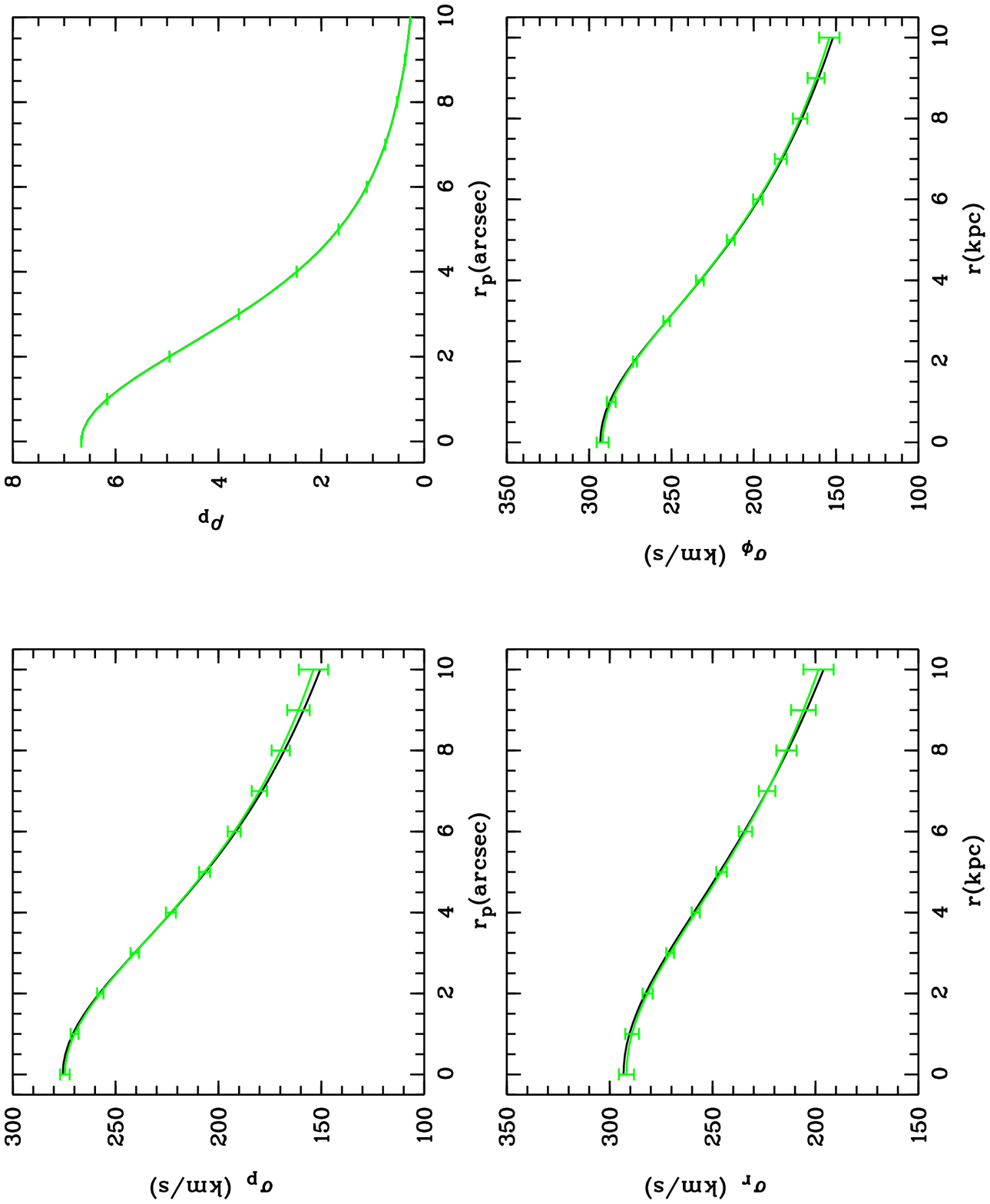"}
\caption{The same as in Fig. \ref{fig2}, but now for a radial
$q=1$ Plummer galaxy. There are no active constraints. \label{fig8}}
\end{figure*}

\section{Conclusion}

As a conclusion we can state that the method can successfully perform
the inversion of equation (\ref{1}), given spectra of high enough
quality (central signal to noise ratio equal to about 50 or higher)
and the correct potential. The errorbars that can be derived from the
Hessian matrix have a statistical meaning and give a good idea of the
uncertainty on the fitted model.

\setcounter{figure}{13}
\begin{figure*} 
\vspace{8cm} 
\special{hscale=52 vscale=52 hsize=570 vsize=230 
	 hoffset=-13 voffset=270 angle=-90 psfile="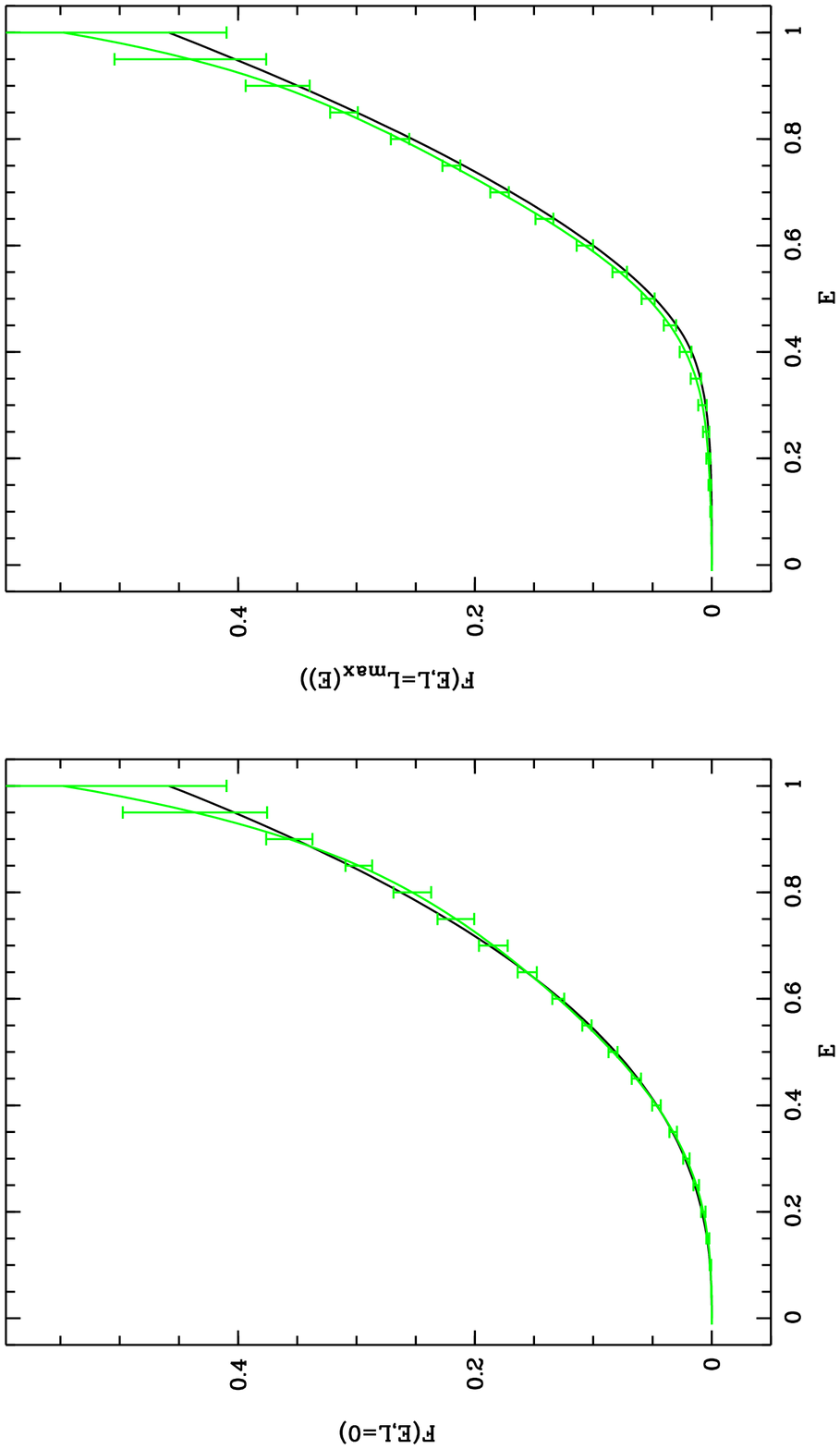"}
\caption{The same as Fig. \ref{fig10} but for a radial $q=1$ 
Plummer galaxy. There are no active constraints. The data used in the fit 
have maximum $S/N \approx 80$. \label{fig40}}
\end{figure*} 

This feature is probably the most important advantage of the present
method. Up to now questions about the reliability of the inferred
dynamics or the adopted potential have remained largely unanswered.
With the present method, there is a link with the $S/N$ of the data,
the choice of the template spectra and the continuum subtraction,
which may all have critical repercussions on the inferred models.
Moreover, some simple form of population synthesis may be attempted
for the best constrained models and geometries.

In recent years, a lot of researchers have addressed the important
subject of constraining the potential of invisible matter (i.e. dark
halo matter, central black holes), using kinematical data (e.g. Merritt
\& Oh \cite{merritt:oh} (who use velocity dispersion data), Rix et
al. \cite{rix:dezeeuw:carollo:vdmarel} (who use $h_3$ and $h_4$ in
addition to projected streaming velocities and dispersions) and
Gerhard et al. \cite{gerhard:jeske:saglia:bender} (who use $h_4$ and
velocity dispersions)). It is not surprising that the inclusion of
LOSVD shape parameters such as $h_3$ and $h_4$ in a modeling
procedure, in addition to projected streaming velocities and
dispersions, gives a significant improvement in determining the
potential. Since in the present method one uses in fact all the
information on the shape of the LOSVDs that can be learned from the
spectra, it should be a valuable tool to constrain the shape and
extent of the potential. 

\appendix

\section{The anisotropic velocity moments}

The anisotropic velocity moments 
\begin{equation} \mu_{2n,2m}(r) = 2
\pi \int \!\! \int F(E,L) v_r^{2n} v_T^{2m} \, dv_r \, dv_T,
\end{equation} 
with $L= r v_T$ and $E = \psi(r) - \frac{1}{2}(v_r^2+v_T^2)$ where
$v_T = \sqrt{ v_\phi^2 + v_\theta^2}$, are connected with the true
velocity moments
\begin{equation} 
\mu_{2n,2i,2j}(r) = \int \!\! \int \!\! \int F(E,L)
v_r^{2n} v_\phi^{2i} v_\theta^{2j} \,dv_r \, dv_\phi \, dv_\theta
\end{equation} 
through the relation 
\begin{equation} \mu_{2n,2i,2j}(r)
= \frac{1}{\pi} B(i+ \frac{1}{2},j+\frac{1}{2}) \mu_{2n,2(i+j)}(r).
\end{equation} 
It can be shown (Dejonghe 1986) that 
\begin{eqnarray}
\mu_{2n,2m}(r) &=& \frac{2^{n+m}}{\sqrt{\pi}}
\frac{\Gamma(n+\frac{1}{2})}{\Gamma(n+m)} \times \nonumber \\ &&
\hspace{-1cm} \int_0^{\psi(r)} (\psi(r)-\psi')^{n+m-1} D_{r^2}^m
(r^{2m} \tilde{\rho}(\psi',r) ) \,d\psi' 
\end{eqnarray} 
with $\tilde{\rho}(\psi,r)$ the augmented mass density, i.e. the density
expressed as a function of $r$ and $\psi$, the potential. Inserting
(\ref{massd}) for the mass density one finds 
\begin{eqnarray}
\mu_{2n,2m}(r) &=& \frac{2^{n+m}}{\sqrt{\pi}} \psi(r)^{\alpha+n+m}
\frac{\Gamma(n+\frac{1}{2}) \Gamma(\alpha+1)}{\Gamma(\alpha+n+m+1)}
\times \nonumber \\ && D_x^m \left(
\frac{x^{m+\beta}}{(1+x)^{\beta+\gamma}} \right) 
\end{eqnarray} 
with $x=(r/c)^2$. Since 
\begin{eqnarray} 
D_x^m \left( \frac{x^{m+\beta}}{(1+x)^{\beta+\gamma}} \right) &=& 
(-1)^m \frac{x^\beta}{(1+x)^{\beta+\gamma}} \times \nonumber \\ 
&& \hspace{-3.5cm} \sum_{j=0}^m \left( \begin{array}{c}
        m \\
        j 
       \end{array} \right) (-m-\beta)_{m-j} (\beta+\gamma)_j 
\left( \frac{x}{1+x} \right)^j \nonumber \\
&=& \frac{\Gamma(m+\beta+1)}{\Gamma(\beta+1)} 
\frac{(r/c)^{2\beta}}{(1+(r/c)^2)^{\beta+\gamma}} \times \nonumber \\
&& \hspace{-3.5cm}
\,_2F_1(-m,\gamma+\beta;1+\beta; \frac{(r/c)^2}{1+(r/c)^2})
\end{eqnarray}
the expression for the velocity moments can be rearranged as 
follows
\begin{eqnarray}
\mu_{2n,2m}(r) &=& \frac{2^{n+m}}{\sqrt{\pi}} 
\frac{\Gamma(n+\frac{1}{2}) \Gamma(\alpha+1) \Gamma(\beta+m+1)}
{\Gamma(\beta+1) \Gamma(\alpha+n+m+1)}  \times \nonumber \\
&& \psi(r)^{\alpha+n+m} \frac{(r/c)^{2\beta}}{(1+(r/c)^2)^{\beta+\gamma}} 
\times \nonumber \\
&& \,_2F_1(-m,\gamma+\beta;1+\beta; \frac{(r/c)^2}{1+(r/c)^2}).
\end{eqnarray}

\section{The LOSVDs}

To calculate the LOSVDs we make use of the fact that the Laplace
transformation of a function $f(x)$ can be written as a power series
containing the moments of that function $\mu_n = \int_0^\infty x^n
f(x)\,dx$ :
\begin{equation}
{\cal L}_{x \rightarrow s} \{ f \} = \sum_{n=0}^\infty \frac{\mu_n}{n!} 
(-s)^n.
\end{equation}
The aim is to calculate $lp(r,r_p,v_p)$, the probability of finding
a star with projected velocity $v_p$ at a distance $r$ from the center
of the cluster and on a line-of-sight with distance $r_p$ to the
center of the galaxy, which has only non-zero even moments
\begin{eqnarray}
\mu_{2n}(r,r_p) &=& \rho(r) \langle v_p^{2n}(r,r_p) \rangle \nonumber \\
&=& \sum_{i=0}^n \left( \begin{array}{c}
                         2n \\
                         2i 
                        \end{array} \right) (\cos \eta)^{2(n-i)} 
(\sin \eta)^{2i} \times \nonumber \\
&& \mu_{2(n-i),2i,0}(r)
\end{eqnarray}
with $\eta(r,r_p)$ the angle between the spherical radial direction
and the line-of-sight at a point of the line-of-sight. Using the above
derived expressions for the velocity moments this becomes
\begin{eqnarray}
\mu_{2n}(r,r_p) &=& \frac{2^n}{\sqrt{\pi}} 
\frac{\Gamma(n+\frac{1}{2}) \Gamma(\alpha+1)}{\Gamma(n+\alpha+1)}
\frac{(r/c)^{2\beta}}{(1+(r/c)^2)^{\beta+\gamma}} 
\times \nonumber \\
&&  \psi(r)^{n+\alpha}
\sum_{i=0}^n \left( \begin{array}{c}
                         n \\
                         i
                    \end{array} \right) (-1)^i (\sin \eta)^{2i}
\times \nonumber \\
&& \sum_{j=0}^n \frac{(-\beta)_{i-j} (\beta+\gamma)_j}{j! (i-j)!} \xi^j
\end{eqnarray}
with $\xi = (r/c)^2/(1+(r/c)^2)$. The Laplace transformation of
$lp(r,r_p,v_p^2)$, for which we use the notation ${\cal L}$, can be
written as
\begin{eqnarray}
{\cal L} &=& {\cal Q} \sum_{n=0}^\infty \frac{\Gamma(n+\frac{1}{2})}
{\Gamma(n+\alpha+1)} \frac{(-2 \psi s)^n}{n!} \sum_{i=0}^n 
\left( \begin{array}{c}
        n \\
        i
       \end{array} \right) \times \nonumber \\
&& (-1)^i (\sin \eta)^{2i} \sum_{j=0}^i 
\frac{(-\beta)_{i-j} (\beta+\gamma)_j}{j! (i-j)!} \xi^j \nonumber \\
&=& {\cal Q} \sum_{j=0}^\infty (\beta+\gamma)_j \frac{\xi^j}{j!} 
(\sin \eta)^{2j} \sum_{i=0}^\infty (-1)^{i+j} \frac{(-\beta)_i}{i!}
\times \nonumber \\
&& \frac{(\sin \eta)^{2i}}{(i+j)!} \frac{\Gamma(i+j+\frac{1}{2})}
{\Gamma(i+j+\alpha+1)} (-2 \psi s)^{i+j}  \times \nonumber \\
&& \,_1F_1 (i+j+\frac{1}{2};i+j+\alpha+1;-2 \psi s).
\end{eqnarray}
with
\begin{equation}
{\cal Q} = \frac{\Gamma(\alpha+1)}{\sqrt{\pi}} \psi(r)^\alpha
\frac{(r/c)^{2\beta}}{(1+(r/c)^2)^{\beta+\gamma}}.
\end{equation}
Making use of the fact that 
\begin{eqnarray}
{\cal L}^{-1}_{y \rightarrow t} \{ _1F_1 (a;b;-y) \} \nonumber &&\\ 
&& \hspace{-3cm} = \frac{\Gamma(b)}{\Gamma(a) \Gamma(b-a)} 
(1-t)^{b-a-1} t^{a-1} 
\hspace{.5cm} 0 < t < 1 \\
&& \hspace{-3cm} = 0 \hspace{4cm} t>1
\end{eqnarray}
and
\begin{eqnarray}
{\cal L}^{-1}_{y \rightarrow t} \{ y^k {\cal L}_{t \rightarrow y} 
\{ f \} \} &=& f^{(k)}(t) + \nonumber \\
&& 
\sum_{n=1}^k f^{(k-n)}(0) {\cal L}^{-1}_{y \rightarrow t} \{ y^{n-1} \},
\end{eqnarray}
where $f^{(k)}$ stands for the $k$-th order derivative, 
one can invert the Laplace transformation. The function $lp(r,r_p,v_p)$ 
can be found by making the connection 
\begin{equation}
lp(r,r_p,v_p) \Leftrightarrow | v_p | lp(r,r_p,v_p^2).
\end{equation}
The final result is 
\begin{eqnarray}
lp(r,r_p,v_p) &=& \frac{1}{\sqrt{2 \pi}} 
\frac{\Gamma(\alpha+1)}{\Gamma(\alpha+\frac{1}{2})} 
\frac{(r/c)^{2\beta}}{(1+(r/c)^2)^{\beta+\gamma}} \times \nonumber \\
&& \hspace{-2cm}
\psi(r)^{\alpha-1/2} \sum_{j=0}^\infty (\beta+\gamma)_j 
\frac{\xi^j}{j!} (\sin \eta)^{2j} \times \nonumber \\
&& \hspace{-2cm}
\sum_{i=0}^\infty 
\left( \begin{array}{c}
        1 \\
        2
\end{array} \right)_{i+j} \frac{(-\beta)_i}{(i+j)!} 
\left( 1- \frac{v_p^2}{2 \psi(r)} \right)^{-(i+j)+\alpha-1/2} \times \nonumber \\
&& \hspace{-2cm}
\frac{(\sin \eta)^{2i}}{i!}\,_2F_1 (-(i+j),\alpha;\frac{1}{2};\frac{v_p^2}{2 \psi(r)}).
\end{eqnarray}
This formal result contains two infinite sums that diverge when
implemented. Only when $\beta$ is a positive integer and
$\beta+\gamma$ is a negative integer do both sums terminate after a
finite number of terms. 

\section{Error analysis on the coefficients}

The minimisation problem is a least squares problem with
constraints. The $r$ active constraints can be dealt with by the
introduction of $r$ Lagrange multiplicators that form a vector
${\rm d}$. The function that has to be minimised is 
\begin{equation}
\chi^2 = {\rm c}^t {\rm{\bf D}} {\rm c} - 2 {\rm p}^t {\rm c} + e + 2 {\rm
d} ^t (\rm{\bf A} {\rm c} - {\rm b}).
\end{equation}

Here the element $A_{is}$ of the matrix $\rm{\bf A}$ is the value of the DF of
the $i$-th component on the $s$-th grid-point in phase space, where the
DF may reach its lower boundary $b^s$, which is usually zero. Standard theory 
leads to the $m$ so-called normal equations
\begin{equation} 
\rm{\bf D} {\rm c} + \rm{\bf A}^t {\rm d} = {\rm p}
\end{equation} 
that have to be solved, along with the $r$ active constraints
\begin{equation}
\rm{\bf A} {\rm c} = {\rm b},
\end{equation}
for the $m$+$r$ unknowns ${\rm c}$ and ${\rm d}$. One obtains the system
\begin{equation}
\rm{\bf M} \left( \begin{array}{c}
          {\rm c} \\
          {\rm d}
         \end{array} \right) 
= \left( \begin{array}{cc}
        \rm{\bf D}  & \rm{\bf A}^t \\
        \rm{\bf A} & \rm{\bf 0} 
       \end{array} \right) \left( \begin{array}{c}
                                  {\rm c} \\
                                  {\rm d} 
                                 \end{array} \right) 
= \left( \begin{array}{c}
          {\rm p} \\
          {\rm b}
         \end{array} \right).
\end{equation}
This can be solved for the coefficients $c_i$
\begin{eqnarray} 
c_i &=& M^{-1}_{ij} p^j + M^{-1}_{is} b^s \nonumber \\
    &=& M^{-1}_{ij} g^j_n g_n w^n + M^{-1}_{is} b^s \nonumber \\
    &=& F_i^n g_n + M^{-1}_{is} b^s. 
\end{eqnarray}
Here $i$ and $j$ range from 1 to $m$ and $s$ varies between 1 and
$r$. The coefficients are written as a linear combination of the data
points with the aid of the matrix $F_i^n= w^n M^{-1}_{ij} g^j_n$ (no
summation over $n$), where $n = 1, \ldots ,N$, which is independent of
the data points $g_n$. The covariance matrix of the coefficients,
$\Sigma_{ij}(c)$, can be found using
\begin{eqnarray} 
\Sigma_{ij}(c) 
&=& \langle (c_i - \bar{c}_i)(c_j-\bar{c}_j) \rangle \nonumber \\
&=& F_i^n F_j^m \langle (g_n - \bar{g}_n)(g_m - \bar{g}_m) \rangle \nonumber \\
&=& F_i^n \Sigma_{nm}(g) F_j^m. 
\end{eqnarray}
The covariance matrix of the data points is approximated by
$\Sigma_{nm}(g) = \delta_{nm}/w^n$. Inserting the explicit 
form of the elements of $\rm{\bf F}$ one finds
\begin{equation}
\Sigma_{ij}(c) = M^{-1}_{il} D^{kl} M_{jk}^{-1}.
\end{equation}
In the absence of positivity constraints on the DF ($\rm{\bf A}=\rm{\bf 0}$) the
covariance matrix of the coefficients is simply the inverse of the
Hessian matrix. Taking the constraints into account is expected to
decrease the width of the errorbars since the positivity constraint
rejects a lot of otherwise acceptable (but unphysical) fits to the
data.

Knowing the error on the coefficients it is also possible to obtain
errorbars on every function that depends on these coefficients
\begin{equation}
\sigma(f(c_1, \ldots,c_m)) \approx \sqrt{ \frac{\partial f}
{\partial c_i} \Sigma_{ij}(c) \frac{\partial f}{\partial c_j}.}
\end{equation}
For instance, the error on a LOSVD of the fitted model becomes 
\begin{eqnarray}
\sigma(\phi(v_p,x_p,y_p)) &=& \nonumber \\
&& \hspace{-2cm} \sqrt{ \phi^i(v_p,x_p,y_p)
\Sigma_{ij}(c) \phi^j(v_p,x_p,y_p)}.
\end{eqnarray}
The elements of the Hessian matrix scale approximately proportional to
the square of the signal to noise ratio. This means that the error on
the coefficients, neglecting the influence of the positivity
constraints, scales inversely proportional to the signal to noise
ratio.

The Hessian matrix is also influenced by the fact whether or not the
components are linearly independent. When two components are linearly
dependent, their coefficients can have an infinity of values without
changing the outcome of the fit, balancing the one component with the
other. The errorbars on the derived quantities, e.g. the LOSVDs, do
not take into account this balancing and just measure the extreme
values of these quantities, given the extreme values of the
coefficients. Consequently, the errorbars will be infinitely large
when linearly dependent components are used. This is reflected by the
Hessian matrix being singular, rendering its inversion
impossible. When nearly linearly dependent components are used, the
Hessian will be close to singular, resulting in large errorbars.

\end{document}